\documentclass[sigconf,preprint]{acmart}
\usepackage{adjustbox}

\usepackage{xspace}

\usepackage{multirow}

\usepackage{enumitem}
\setlist[enumerate]{leftmargin=*}
\setlist[itemize]{leftmargin=*}
\usepackage{caption}

\renewcommand{\paragraph}[1]{\smallskip\noindent\textbf{#1.\mbox{\ \ }}}

\AtBeginDocument{%
  \providecommand\BibTeX{{%
    \normalfont B\kern-0.5em{\scshape i\kern-0.25em b}\kern-0.8em\TeX}}}

\setcopyright{none}
\acmConference[Accepted to WWW'23]{}{Austin}
\begin{document}
\newcommand{\sg}[1]{{\color{blue}SG: #1}}
\newcommand{\sr}[1]{{\color{orange}SR: #1}}
\newcommand{\GW}[1]{{\color{red}GW: #1}}
\newcommand{\edit}[1]{{\color{black}#1}}

\newcommand{\numclasses}{90\xspace}
\newcommand{\numdomains}{6\xspace}
\title{Class Cardinality Comparison as a Fermi Problem}

\author{Shrestha Ghosh}
\affiliation{%
  \institution{Max Planck Institute for Informatics}
  \institution{Saarland University}
  \city{Saarbruecken}
  \country{Germany}
}
\email{ghoshs@mpi-inf.mpg.de}

\author{Simon Razniewski}
\affiliation{%
  \institution{Max Planck Institute for Informatics}
  \city{Saarbruecken}
  \country{Germany}
}
\email{srazniew@mpi-inf.mpg.de}

\author{Gerhard Weikum}
\affiliation{%
  \institution{Max Planck Institute for Informatics}
  \city{Saarbruecken}
  \country{Germany}
}
\email{weikum@mpi-inf.mpg.de}

\renewcommand{\shortauthors}{Ghosh et al.}

\begin{abstract}

Questions on class cardinality comparisons are quite tricky to answer and come with its own challenges.   
They require some kind of reasoning since web documents and knowledge bases, indispensable sources of information, rarely store direct answers to questions, such as, ``Are there more astronauts or Physics Nobel Laureates?'' 
We tackle questions on class cardinality comparison by tapping into three sources for absolute cardinalities as well as the cardinalities of orthogonal subgroups of the classes. We propose novel techniques for aggregating signals with partial coverage for more reliable estimates and evaluate them on a dataset of 4005 class pairs, achieving an accuracy of 83.7\%.
\end{abstract}

%    <concept>
%        <concept_id>10010147.10010178.10010179.10003352</concept_id>
%        <concept_desc>Computing methodologies~Information extraction</concept_desc>
%        <concept_significance>500</concept_significance>
%        </concept>
%    <concept>
%        <concept_id>10010147.10010178.10010187.10010188</concept_id>
%        <concept_desc>Computing methodologies~Semantic networks</concept_desc>
%        <concept_significance>500</concept_significance>
%        </concept>
%  </ccs2012>

% \ccsdesc[500]{Computing methodologies~Information extraction}
% \ccsdesc[500]{Computing methodologies~Semantic networks}
\keywords{Entity classes, Comparative questions, Class cardinality estimation}



\maketitle

\section{Introduction}
\paragraph{Motivation and Problem}
Are there more astronauts or Physics Nobel laureates? More nuclear power plants or catholic cathedrals? 
More lakes or rivers?
More airports on this planet than satellites in orbit, or vice versa?
Comparative questions of this kind tickle our curiosity, yet are often surprisingly hard to answer. 
For some comparisons, there are authoritative official sources that provide reliable (albeit not necessarily up-to-date) numbers.
In most cases, though, the natural resort is
to tap into online sources like knowledge bases (KBs, e.g., Wikidata), search engine (SE, e.g., Bing) results, or large language models (LMs, e.g., GPT).
However, all of these come with biases in what they cover and what not, and often give treacherous signals that lead humans to wrong conclusions.

For example, the Wikidata knowledge base
suggests that there are more than 400,000 rivers and roughly 23,000 castles.
While we found the number of rivers to be close to 300,000~\cite{grill2019mapping}, the number of castles is quite difficult to obtain and lies for Europe alone perhaps between 400,000 to 1.3 million\footnote{\url{https://www.quora.com/How-many-castles-are-there-in-Europe/answer/Michael-Burke-339}; \url{https://www.dw.com/en/does-germany-really-have-25000-castles/a-42350502}}.

Major search engines, when probed with different query formulations, pick up on the number of rivers as 250,000 (in the United States) and 10,000 (medieval European) castles, suggesting that there are far more rivers than castles, contradicting reality.
Even the GPT-3 language model, which performs well on question answering tasks, is completely off. When prompted with three example questions, followed by the question asking for the number of rivers, it returns \textit{``an estimated 1.3 million rivers on Earth''} and for castles, it returns, \textit{``There are around 900 castles in the world.''}
Obviously, neither of these major sources is a true mirror of reality; the online world is inherently hampered by incompleteness and bias.

Smart humans, on the other hand, are sometimes able to judiciously select online sources as cues, then combine multiple cues in a clever way, and eventually arrive at reasonable estimates at class cardinalities (at least the right order of magnitude) and relative comparisons between classes (just asserting which is bigger). 
Enrico Fermi, a Physics Nobel Laureate from the early 20th century, was known to be a master of such estimates; hence this kind of problem is also known as Fermi Problem
\cite{fermi}. 
In this paper, we aim to emulate a smart human's approach. We introduce and study a variety of online signals that could be brought to bear, gaining insights on strengths and weaknesses for different domains of entity classes (e.g., occupations vs. creative works vs. man-made physical objects).  Moreover, we propose novel techniques for aggregating signals with partial coverage into more reliable estimates on which of the two given classes has more real-world instances.

\paragraph{{Approach and Contribution}} 
This paper focuses on {\em dominance estimation}: which of two classes has the higher cardinality. We obtain cues for the numeric cardinalities from three sources: the Wikidata KB via SPARQL queries~\cite{vrandevcic2014wikidata}, the Bing search engine with judicious queries using the CoQEx method \cite{Ghosh2022answeringb}, and the GPT-3 language model \cite{brown2020language} with various prompts.
Absolute cardinalities from these sources are often completely wrong; so we interpret them merely as signals to be used for further inference. 

%GW: now one par on the paper's key idea: aggregation over subgroups
The key idea of mitigating these bias effects is to additionally inspect {\em subgroups} of classes, such as actors or airports by country or geo-region (e.g., North America, East Asia etc.). 
Such subgroups are orthogonal to the classes under comparison. The hypothesis that we study is that the estimates for subgroups of classes can give more reliable cardinality estimates, at least for some subgroups and for relative comparison.
This larger set of finer-grained signals are then aggregated using different techniques proposed in this paper.

The evaluation dataset consists of 4005 class pairs from 6 diverse domains.
Our major finding is that the novel technique of aggregating signals substantially improves the dominance estimation, achieving over 80\% accuracy compared to direct source signals.

\section{Related Work}\label{sec:related_work}
We focus on bias in the digital world and completeness in information sources, which would most affect dominance estimation.

\paragraph{\textbf{Bias in the Digital World}}
Wikipedia, a major source of general knowledge, used in automatic construction of KBs, suffers from implicit and explicit bias~\cite{Hube2017BiasIW} and specific demographic biases such as gender bias~\cite{sun2021men}. 
The common crawl \cite{commoncrawl}
is another massive dataset of textual information, almost 50 times greater than Wikipedia (around 5.6 TB) and the BookCorpus (6 GB). It requires significant processing before it can be used for pre-training large LMs~\cite{raffel2020exploring}. A case study shows that the filtered C4 dataset disproportionately affects minorities~\cite{Dodge2021DocumentingLW}. Another work addresses the different biases on the web itself and its potential effects~\cite{BaezaYates2018BiasOT}.
Representation in KBs is unbalanced 
due to multiple reasons, including but not limited to reporting bias~\cite{gordon2013reporting} as well as data, schema and inferential bias~\cite{Janowicz2018DebiasingKG,safavi2021report}.

\paragraph{Sources of Information}
It is well known that general-knowledge KBs 
are incomplete~\cite{Razniewski2016ButWD,Weikum2021MachineKC} even with their increased coverage over time~\cite{Razniewski2020StructuredKH}. 
Species estimation techniques from biology have been used to estimate cardinalities \cite{Trushkowsky2013CrowdsourcedEQ, luggen2019non}.
Techniques like mark and recapture assume sampling from the real-world and, applying them to KB edit history makes severe underestimations, for instance predicting that there are roughly 4M humans \cite{luggen2019non}\footnote{{\url{https://cardinal.exascale.info/}}}. 
Current powerful SEs 
are now capable of providing structured answers from their internal KBs, and highlighting the most probable answer in the top snippet. Even so, structured answers are not the norm
for more complex or less popular questions. This low recall has prompted research on answering entity counts from multiple SE snippets~\cite{Ghosh2022answeringb,Ghosh2022answeringa} and large-scale mining of quantities from the web~\cite{Elazar2019HowLA}. 
LMs have been shown to be effective in recalling factual information~\cite{karpukhin2020dense,petroni2019language}. Nevertheless, they lack scrutability, rely on high quality prompts,
~\cite{Jiang2020HowCW}, and are known to struggle with numeric/count information~\cite{lin2020birds}.

\section{Class Cardinality Comparison}
Let us continue with the question: \textit{Are there more rivers than castles?} In principle, such questions can be decomposed into cardinality questions: \textit{how many rivers are there?} and \textit{how many castles are there?}, the answers of which are then compared.
% \subsection{Dominance Estimation}
We therefore identify three problem statements related to comparison questions, of inversely related informativeness and difficulty.
\begin{enumerate}
\item[1.] \textit{Cardinality Estimation}: What are the absolute cardinalities of the classes \textit{rivers} and \textit{castles}?
\item[2.] \textit{Proportionality Estimation}: What is their ratio?
\item[3.] \textit{Dominance Estimation}: Are there more \textit{rivers} or \textit{castles}?
\end{enumerate}
While cardinality estimation provides the most information,
this is also the most challenging task. 
Estimation methods may give impractical results that are orders of magnitude away from the ground-truth, and for some classes, there might not even exist a widely agreed ground-truth.

Looking at the relation of pairs of classes reduces the impact of uncertainty: Even if there is no widely agreed count for the number of castles, most estimates might agree that there are more castles than rivers. Determining the actual proportion would be desirable, though it is also subject to uncertainty.

In this paper, we focus on the most approachable task of \textit{dominance estimation}: to determine whether one class is bigger than the other. 
Intuitively, as humans, we can deduce that there are more castles than rivers from our observation that along a river, there are typically several castles. However, \textit{machines are incapable of such reasoning}. When the classes do not come with such common observations, such as comparing the number of rivers and satellites, the task becomes difficult even for humans. Nevertheless, there exists evidence in the form of KB entities and actual counts in web documents, which can be leveraged to predict the bigger of the two classes. 
We identify cardinality signals to predict whether a class $A$ is greater than $B$, and define the output variable as:
\begin{align}
O_{|A|>|B|} = 
\begin{cases}
    1 & \text{if $|A|$ is predicted to be bigger than $|B|$}, \\
    -1 & \text{if $|A|$ is predicted to be smaller than $|B|$}, \\
    0 & \text{if the predictor abstains.}
\end{cases}
\end{align}

\subsection{Basic Cardinality Signals}

We obtain cardinality signals from three different sources, and of two types.

\paragraph{Signal sources}
We use three different signal sources: KBs, SEs, and LMs. Each provides a different angle: of entities covered in a KB, of what is popular on the web, and of what information has been distilled by LMs. Specifically we look into Wikidata~\cite{vrandevcic2014wikidata}, the top-50 search results by Bing~\cite{bing} and GPT-3~\cite{brown2020language}.

\begin{enumerate}
\item[1.] {\textbf{Knowledge base (KB).}}
Here we formulate SPARQL queries to retrieve the count of entities per class from Wikidata,
preferring hand-annotated queries over KB-QA systems. KB-QA systems perform worse than hand-annotated queries due to the available system's difficulty to formulate the most accurate SPARQL queries.
\item[2.] {\textbf{Search engine (SE).}} Return the most confident cardinality of a class using the CoQEx system for inferring counts from top-50 SE result snippets~\cite{Ghosh2022answeringa}.

\item[3.] {\textbf{Language model (LM).}} In a few-shot setting, given $n$ cardinality questions and their answers as a prompt, retrieve the cardinality of the class in the $(n+1)^{th}$ question.
Specifically, we provide three examples along with the intended question to the GPT-3 model. The examples and the model parameters remain constant for all queries. 
We process the text output to extract the cardinality of the class using Python Quantulum3 library\footnote{{\url{https://pypi.org/project/quantulum3/}}}. 
\end{enumerate}

\paragraph{Signal types}
We consider two types of signals.
\begin{enumerate}
    \item[1.] \textbf{Root signals}. Signals for the count of the class of interest itself, e.g., \textit{rivers} (worldwide).
    \item[2.] \textbf{Subgroup signals}. These are signals for counts of subgroups of the group of interest, e.g., \textit{castles in Germany}.
\end{enumerate}

\edit{What are orthogonal subgroups and how to select them?}

In principle, subgroups signals can be computed for a range of subgroups (e.g., 195 countries, by year/decade, by status, etc.). In the following we focus on the G20-group of countries \cite{g20}, as for these, sources tend to have more reliable information.

Note also that subgroup signals normally \textit{do not sum up} to root signals, both for pragmatic reasons (data for certain subgroups is unavailable/incomplete), as well as principled reasons (an entity belonging to several subgroups, or to none).

\subsection{Signal Aggregation}
\edit{Besides using the root signals directly, we aggregate basic signals to predict class comparison.}
We propose to proceed in three levels: First we aggregate subgroup signals, after which we include root signals, and then aggregate the resulting signals by sources.

\paragraph{I. Subgroup aggregation}
\begin{enumerate}
    \item[1.] By majority: $O_{|A|>|B|}^\textit{M}$ is 1, if at least $\theta_M$ percent of the subgroups of A have more entities than the those of B, -1 if at least $\theta_M$ have less, else 0. 
    \item[2.] By significance: We perform a one-sided t-test, testing for A bigger than B when the mean of the subgroup cardinalities of A is greater than that of B. We test for A less than B when the reverse is true. In the former case, if the p-value $\leq \alpha$, the subgroup distribution of A is significantly greater than B and $O_{|A|>|B|}^\textit{S}$ is 1. If the p-value $\leq \alpha$ in the latter case, then the reverse is true and $O_{|A|>|B|}^\textit{S}$ is -1.
    If the p-value $> \alpha$ in either of the cases, the prediction is 0. 
\end{enumerate}

\paragraph{II. Root and subgroup aggregation}
Here, the final prediction is the average over the predictions obtained from i) comparing root signals, and ii) majority vote over subgroup signals, and iii) significance test over subgroup signals. We use weights, $W \in [0,1]$ for the majority and significance predictions, such that, 
\begin{equation}
    \textit{Ensemble}_{|A|>|B|}^\textit{source} = 
        \frac{1}{3} (O_{|A|>|B|}^\textit{source} + W_\textit{M} \cdot O_{|A|>|B|}^\textit{source,M} + W_\textit{S} \cdot O_{|A|>|B|}^\textit{source,S})
\end{equation}
Hereby, the weight $W_M$ is the majority ratio, when majority is $>\theta_1$, \textit{i.e.} A bigger than B, and, $W_M =$ $(1-\textit{majority ratio})$ when B is greater. The significance prediction is weighted by (1-p-value), such that the lower the p-value, higher the significance.

\paragraph{III. Final source aggregation}
In the last level of aggregation, we combine the predictions from all sources by:

\noindent \textit{Majority vote}: if any two sources agree, that label is selected, else the predictor abstains.

\noindent\textit{Weighted vote}: we train a Logistic Regression classifier for each of the signal aggregation cases, to learn the weights of each source. 
\begin{align}
    ln(p(|A|>|B|)) &= -ln(1+ e^{-\sum_{\textit{source}} W_\textit{source} \cdot \textit{Ensemble}_{|A|>|B|}^\textit{source} })\\
    \textit{Ensemble}_{|A|>|B|}^{\textit{weighted}} &=
    \begin{cases}
        1 & \text{if p(|A|>|B|) > p(|A|<|B|)}, \\
        -1 & \text{if p(|A|>|B|) < p(|A|<|B|)}, \\
        0 & \text{otherwise.}
    \end{cases}
    % \sum_{\textit{source}} W_\textit{source} \cdot \textit{Ensemble}_{|A|>|B|}^\textit{source} 
\end{align}

\section{Experiment}

\paragraph{{Dataset}}
We create a ground-truth (GT) dataset of \numclasses classes spanning \numdomains domains (Table~\ref{tab:domains})\footnote{Dataset link: \url{https://github.com/ghoshs/class_cardinality_comparison}.}. 
Our evaluation set comprises $\binom{90}{2} = 4005$ combinations of class pairs, of which $6 \times \binom{15}{2} = 630$ are in-domain pairs, i.e., both classes belong to the same domain and the remaining 3375 class pairs are inter-domain, i.e., both classes belong to different domains. In order to assess the difficulty of the task, we compute the order of magnitude of the ratio of the ground-truth cardinalities of all class pairs. We argue that class pairs with close cardinalities would be more difficult to predict than class pairs whose cardinalities differ by several orders of magnitude.
The dataset has more pairs with close cardinalities, and less than 8\% of the class pairs have cardinality ratios more than $10^4$ orders of magnitude. For instance, airlines (5,000) and national parks (3,369) has a cardinality ratio of $1.4 : 1$ (higher:lower)
while airlines to politicians (6,500,000) has a cardinality ratio of $1.3 \times 10^3 : 1$.

\begin{table}[t]
    \centering
    \caption{Domains and example classes.}
    \label{tab:domains}
    \begin{adjustbox}{width=0.4\textwidth}
        \begin{tabular}{|l|ll|}
            \hline
            \textbf{Domain} & \textbf{\#Classes} & \textbf{Examples}\\\hline
            creative work & 15 & film, board game, book\\
            geographical entities & 15 & lake, castle, dam\\
            man-made object & 15 & satellite, submarine\\
            occupation & 15 & politician, actor, physicist\\
            organization & 15 & university, football club \\
            species & 15 & snake, insect, fish\\
            \hline
        \end{tabular}
    \end{adjustbox}
\end{table}

\paragraph{{Metrics}}
We primarily measure the performance of each signal by its \textit{accuracy}, i.e., as the percentage of correct predictions relative to all samples. To also see whether low accuracies stem from abstaining often, or wrong predictions, we additionally report the \textit{rate of abstention}, and, \textit{precision}, as the percentage of correct predictions relative to non-abstentions.

\paragraph{{Parameters}} 
We use the \textit{text-curie-001} model of GPT-3 with temperature set to 0 and maximum tokens to generate set to 15. The subgroup aggregation parameters are $\theta_M$, which we set to 0.5 (more than 50\% majority) and $\alpha$, which we set to 0.05. 

For training the Logistic Regression classifiers, we divide the dataset into train and test splits (80:20). We perform a 5-fold cross-validation on the training data to determine the regularization hyperparameter.
We report evaluation on the whole dataset of 4005 class pairs, except for the weighted vote ensemble, which is evaluated on the test set of 800 samples.

\begin{table*}[t]
    \begin{minipage}[t][][t]{0.62\textwidth}
    \centering
    \caption{Accuracy (in \%) of cardinality signals. 
    }
    \label{tab:acc_by_source}
    \begin{adjustbox}{width=\textwidth}
        \begin{tabular}{|l|c|cc|ccc|}
            \hline
            \multirow{2}*{\textbf{Source}} & \multirow{2}*{\textbf{Root (1)}} & \multicolumn{2}{c|}{\textbf{Subgroup Aggregations}} & \multicolumn{3}{c|}{\textbf{Ensemble over Root and Subgroup Signals}}\\
            \cline{3-7}
            & & \textbf{Majority (2)} & \textbf{T-test (3)} & \textbf{(1)+(2)} & \textbf{(1)+(3)} &  \textbf{(1)+(2)+(3)}\\
            \hline
            KB & 64.7 & 57.0 & 36.0 & 61.5 & \textbf{65.9} & 61.8\\
            SE & 65.4 & 67.2 & 39.6 & 65.7 & 65.4 & \textbf{68.4} \\
            LM & 74.4 & 75.8 & 57.3 & 77.1 & 75.8 & \textbf{79.4}\\
            \hline
            \multicolumn{7}{|c|}{\textbf{Ensemble over Sources (KB, SE, LM)}} \\
            \hline
            Majority Vote & 77.8 & 76.3 & 42.9 & 76.8 & 78.7 & \textbf{78.9}\\
            Weighted Vote & 78.2 & 76.2 & 81.2 & 79.3 & \textbf{83.7} & 81.3\\
            \hline
            \multicolumn{7}{|c|}{\textbf{Non-expert human baseline}} \\ 
            \hline
            Closed-book & \multicolumn{6}{c|}{75.0}\\
            Open-book & \multicolumn{6}{c|}{76.0}\\
            \hline
        \end{tabular}
    \end{adjustbox}
    \end{minipage}%
    \hfill
    \begin{minipage}[t][][t]{0.32\textwidth}
    \centering
    \caption{Accuracy (\%) of aggregation over root and subgroups by domain.}
    \label{tab:acc_by_domain}
    \begin{adjustbox}{width=\textwidth}
        \begin{tabular}{|l|llll|}
            \hline
            \textbf{Domain} & \textbf{KB} & \textbf{SE} & \textbf{LM} & \textbf{Best}\\
            \hline
            Creative work & 62.8 & 54.2 & \textbf{83.8} & LM\\
            Geographical entity & \textbf{77.1} & 60.9 & 70.4 & KB\\
            Man-made object & 26.6 & 77.1 & \textbf{96.1} & LM\\
            Occupation & 57.1 & \textbf{80.0} & 74.2 & SE\\
            Organization & 58.0 & 72.3 & \textbf{88.5} & LM\\
            Species & 61.9 & \textbf{78.0} & 63.8 & SE\\
            % \hline
            \textit{Interdomain} & 62.6 & 68.0 & \textbf{79.4} & LM \\
            \hline
            All & 61.8 & 68.4 & \textbf{79.4} & LM \\
            \hline
        \end{tabular}
    \end{adjustbox}
    \end{minipage}
\end{table*}

\paragraph{Human Baseline}
We sampled 100 pairs, \edit{50 in-domain and 50 inter-domain class pairs, to evaluate non-expert human performance, emulating general users who may not be domain experts}. 
Each class pair was annotated by three MTurk annotators, who were given a brief description of each class, and were asked: 
\begin{enumerate}
    \item Which class has more entities in real life? \\
    \textit{Class 1, Class 2, Equal, Cannot determine}
    \item How certain are you? 
    \textit{Very sure, Estimation, Guess, No idea}
\end{enumerate}
In addition, annotators could justify their responses in a free text field.
The task had two settings (i) the closed-book setting, where annotators should answer without consulting external sources, and (ii) the open-book setting, where the annotators were encouraged to perform web research. For the closed-book settings, annotators had 3 minutes per question, for the open-book setting, 8 minutes.

The accuracy in both settings was comparable (75\% vs.\ 76\%), although the precision increased substantially more (from 79\% to 85\%), i.e., in the open-book setting, the additional evidence made annotators more often abstain, instead of guessing.

\edit{\paragraph{State-of-the-art Baselines}
The signal aggregation methods are compared against the state-of-the-art baselines which come from the root signals of the three sources (KB, SE, LM).}

\begin{table*}[t]
\centering
\caption{Examples from our dataset, ordered from easier to harder.}
% \vspace{-0.3cm}
\label{tab:anecdotes}
\begin{adjustbox}{width=\textwidth}
    \begin{tabular}{|lll|cccccccc|l|}
    \hline
    \multirow{2}{*}{\textbf{Class 1}} & \multirow{2}{*}{\textbf{Class 2}} & \multirow{2}{*}{\textbf{GT ratio}} & \multicolumn{3}{c|}{\textbf{Root signals}} & \multicolumn{3}{l|}{\textbf{Basic signal ensembles*}} & \multicolumn{2}{l|}{\textbf{Source ensembles**}} & \multirow{2}{*}{\textbf{Comment}}   \\
    & & & KB & SE & \multicolumn{1}{c|}{LM} & KB & SE       & {LM}            &  \multicolumn{1}{|c}{Majority} & Weighted & \\ \hline
    Websites & Religious texts & $6 \times 10^7 : 1$ & \checkmark & \checkmark & \checkmark & \checkmark & \checkmark & \checkmark & \checkmark & \checkmark & Strong signals from all sources.\\
    Bacteria species & Bee species & $3.15 \times 10^5 : 1$ & \checkmark & x & x & \checkmark & x & x & x & x & KB root signal ratio of 4.8:1 gives correct prediction.\\
    Books & Paintings & $1.3 \times 10^3 : 1$ & x & x & \checkmark & x & x & \checkmark & x & \checkmark & Weighted vote picks up weak signal from LM.\\
    School teachers & Hospitals & $6.07 \times 10^2 : 1$ & x & \checkmark & x & x & \checkmark & \checkmark & \checkmark & \checkmark & Correct signals from subgroup agg. in SE and LM. \\
    Cities & Islands & $9.7:1$ & x & \checkmark & x & x & \checkmark & \checkmark & \checkmark & x & Weighted vote fails due to strong incorrect KB + weak correct LM signals.\\
    Actors & Architects & $1.48 : 1$ & \checkmark & x & x & \checkmark & x & x & x & \checkmark & Weighted vote leverages strong KB signal.\\
    \hline

    \multicolumn{12}{l}{\textit{\textbf{*} Best ensemble for KB: (1)+(3); SE and LM: (1)+(2)+(3)} from basic signals: Root (1), Majority agg. (2) and T-test agg. (3).}\\
    \multicolumn{12}{l}{\textit{\textbf{**} Best source ensemble for majority vote: (1)+(2)+(3); weighted vote: (1)+(3).}}\\
    
    \end{tabular}    
\end{adjustbox}
\end{table*}

\paragraph{Result 1: Performance by Source}
Table~\ref{tab:acc_by_source} shows the accuracy of the sources by their signals. 
Of the three sources, only LM surpasses the human baseline. We find that an ensemble over the root and the aggregated subgroup signals performs well for all three sources. 
Aggregating over subgroup and root signals reduces the rate of abstention from 0.5\% to 5\%, down to less than 1\% consistently across sources. Upon further inspection we find that the t-test subgroup aggregations have the lowest accuracy, despite good precision (LM: 87\%, SE: 83\%), due to very high abstention rates (LM:34\%, SE: 52\%).

\paragraph{Result 2: Aggregation over Sources}
From Table~\ref{tab:acc_by_source}, we see that learning weights using supervised learning increases robustness across aggregation strategies, as the accuracy does not drop below 76\%, hence performing better than any individual source. On the contrary, majority vote over sources is most effective only on root signals.
\textit{Notably}, in the weighted vote strategy, subgroup aggregations by t-test does remarkably well achieving 81\% when aggregating just the t-test signals and 83.7\% when aggregating the root and t-test signals. In both cases, we noticed high coefficient for LM, followed by lower coefficients for SE and then KB. 

\paragraph{Result 3: Performance by Domain}
We analyse the class pairs by each of the 6 domains on the ensemble over root and subgroup signals in Table~\ref{tab:acc_by_domain}. Notice that there is no one winning source and relying on the best source per domain gives an accuracy above 77\%.
Estimates of everyday objects such as bicycles or smartphones, are easily available on the web, as reflected in SE (77\%) and LM (96\%) accuracies. KB is most accurate on geographical entities (77.14\%), outperforming the other two sources by a large margin.   

\paragraph{Discussion}
Direct comparison questions are more challenging for current SEs and LMs.
For instance, Bing returns irrelevant answers to the question \textit{are there more rivers or lakes?}, while GPT-3 return a definitive answer that there are more lakes. When prompted for an explanation, the answer is reversed. 
Table \ref{tab:anecdotes} shows a few examples from our dataset and the predictions of root signals and the best performing ensembles in each row from Table~\ref{tab:acc_by_source}.

\section{Conclusion}
In this work we tackle dominance estimation of two classes in the real-world by using signals from varied information sources. We propose techniques to combine cardinality signals by aggregating over orthogonal subgroups and over multiple sources. 
From experiments, we observe ensembles over sources using subgroup aggregations to perform very well in dominance estimations.
\bibliographystyle{ACM-Reference-Format}
\bibliography{refs}

\end{document}